\documentclass[a4paper]{jpconf}
\usepackage{graphicx}
\usepackage{amsmath}
\usepackage{comment} 
\usepackage{CJKutf8}
\usepackage{epstopdf}
\usepackage[backref]{hyperref} 
\hypersetup{colorlinks,
	linkcolor=blue,%
	citecolor=blue,
        urlcolor =blue}
\usepackage{ulem}
\usepackage{color}

\begin{document}
\begin{CJK}{UTF8}{gbsn}
\title{Sensing and control scheme for the inteferometer configuration with an L-shaped resonator}

\author{Xinyao Guo}
\address{ 
Zhili College, Tsinghua University, Beijing 100084, China}
\author{Teng Zhang}
\address{School of Physics and Astronomy, and Institute for Gravitational Wave Astronomy,University of Birmingham, Edgbaston, Birmingham B15 2TT, UK}
\ead{tzhang@star.sr.bham.ac.uk}
\author{Denis Martynov}
\address{School of Physics and Astronomy, and Institute for Gravitational Wave Astronomy,University of Birmingham, Edgbaston, Birmingham B15 2TT, UK}
\ead{d.martynov@bham.ac.uk}
\author{Haixing Miao}
\address{Department of Physics, Tsinghua University, Beijing 100084, China}
\ead{haixing@tsinghua.edu.cn}

\begin{abstract}

 The detection of high-frequency gravitational waves around kHz is critical to understand the physics of binary neutron star mergers. A new interferometer design has been proposed in [Phys. Rev. X {\bf 13}, 021019 (2023)], featuring an L-shaped optical resonator as the arm cavity, which resonantly enhances kHz gravitational-wave signals. This new configuration has the potential to achieve better high-frequency sensitivity than the dual-recycled Fabry-Perot Michelson. In this paper, we propose a sensing and control scheme for this configuration. Despite having the same number of length degrees of freedom as the dual-recycled Fabry-Perot Michelson, the new configuration requires one less degree of freedom to be controlled owing to {the system's insensitive to the fluctuation of $L_- - l_-$}. We have also shown that introducing the Schnupp asymmetry is ineffective for controlling the signal-recycling cavity length. Therefore, we propose adding control fields from the dark port to control this auxiliary degree of freedom.

\end{abstract}

\section{Introduction}\par

Ground-based gravitational-wave (GW) detectors, which have made breakthrough discoveries, are dual-recycled Fabry-Perot Michelson interferometers,
comprising a Michelson interferometer with each arm formed by a km-scale Fabry-Perot cavity\,\cite{Aasi2015c, Acernese_2015, Abbott2016b, GW170817, GWTC}. In addition, a power recycling cavity\,\cite{Drever_1983,Abramovici_PLA96} and signal recycling cavity\,\cite{Mizuno_PLA1993, Meers_PRD1988} are introduced at the bright port and dark port respectively to enhance the arm cavity power and shape the detector response to GW signals. Motivated by the observation of the binary neutron star merger event GW170817\,\cite{GW170817}, numerous proposals have emerged for modifying the configuration of GW detectors to enhance their sensitivity at kHz frequencies and better capture such events\,\cite{Miao2018, Martynov2019, Ackley2020,Adya2020, Page2021, Zhang2021, Wang2022}.

These kHz detector proposals leverage the Michelson configuration, that employs arm cavities. However, due to the limited bandwidth of the arm cavity, it can only efficiently amplify signals below the bandwidth, while signals in the high-frequency kHz regime remain largely unamplified. Specifically, when the wavelength of the gravitational wave is twice the length of the arm cavity, the optical phase induced by the GW for the first half of the wave period is opposite to that of the second half, resulting in a null response in the first free spectral range of the cavity. The presence of optical loss in the signal-recycling cavity further limits sensitivity at high frequencies\,\cite{Miao2019, Korobko2023}.
Recently, a new configuration with dual-recycled structure and an L-shaped cavity resonator has been proposed\,\cite{new.configuration}, which shares the same core resonator as the Fox-Smith interferometer\,\cite{Fox_Smith, PhysRevD.Fox}. In contrast to the linear cavity, the signal 
is resonant in the first free spectral 
range for the L-shaped cavity, because 
the sign of the 
optical phase is flipped for the second wave period. 

Sensing and control schemes for the new configuration have not been studied so far. In this work, we present the first preliminary design of the control 
scheme, which is based on the sensing and control idea of Advanced LIGO\,\cite{Fritschel_Appl2001,Abbott_ligo2010, Staley_CQG2014, Aasi2015c, Izumi_CQG2017}. The paper is organized as follows: In Sec.\,\ref{sec:overview}, we provide an overview of the sensing and control scheme of the new interferometer; In Sec.\,\ref{sec:opt}, we present a comprehensive analysis of the optical characteristics of the new configuration;
In Sec.\,\ref{sec:transferfunctions}, we derive the sensing matrix of our scheme analytically, and compare it with numerical results using Optickle, demonstrating good matching between them. We also  provide a diagonalization procedure for the system and discuss the feasibility of our locking scheme in the same section.  Finally, we conclude our study in Sec.\,\ref{sec:conclusion}.\par

\section{Overview of the interferometer sensing and control}\label{sec:overview}

\begin{figure}[t!]
\centering
\includegraphics[scale=0.7]{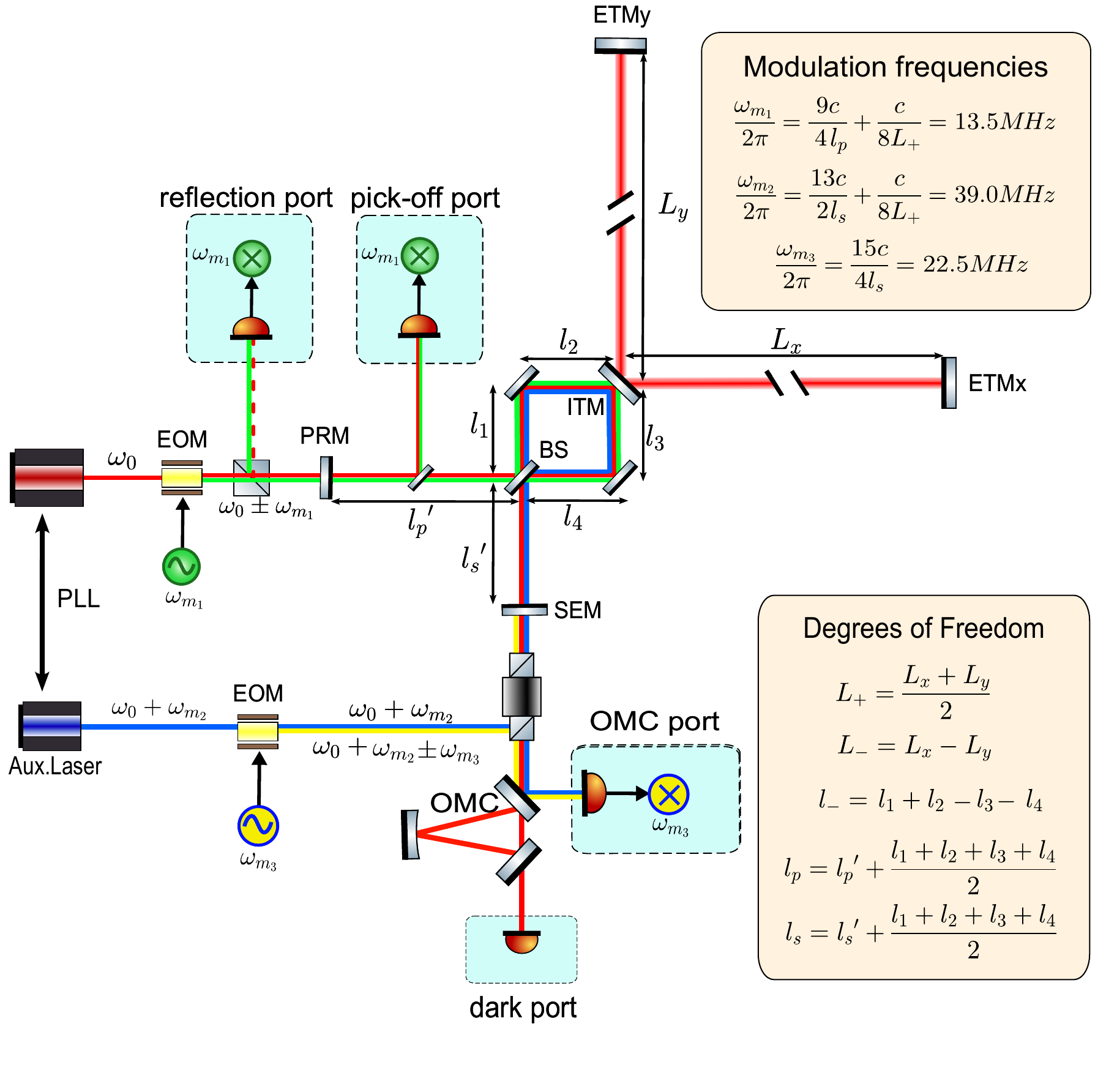}
\caption{A simplified optical layout of the new configuration and illustration of the sensing and control scheme. EOM: electro-optic modulator; PLL: phase locking loop; BS: beam splitter; ITM: input test mass;  ETM: end test mass; PRM: power recycling mirror; SEM: signal extraction mirror; OMC: output mode cleaner}
\label{fig:config}
\end{figure}

In this section, we provide an overview of the sensing and control strategies for the new interfermeter.  
As depicted in Fig.\,\ref{fig:config}, we define five degrees of freedom,
\begin{align}\label{eq121}
L_+&=\frac{L_x+L_y}{2},\\ L_-&=L_x-L_y,\;\\l_-&=l_1+l_2-l_3-l_4, \\ l_p&=l_{p}'+\frac{l_1+l_2+l_3+l_4}{2},\;\\l_s&=l_{s}'+\frac{l_1+l_2+l_3+l_4}{2}\,.
\end{align}
Here, $L_+$ and $L_-$ define the common and differential modes of the L-resonator, representing the average and difference of the arm lengths, $L_x,L_y$, respectively. $l_-$ defines the ``Michelson" mode, referring to the length difference between the upper and lower paths in the central loop. $l_p$ and $l_s$ define the lengths of the power recycling cavity and signal extraction cavity respectively, where $l_{p}'$ and $l_{s}'$ are the distances from the beamsplitter to the power recycling mirror and signal recycling mirror, respectively.

In this study, we propose a sensing and control scheme that involves the incorporation of four open ports and three radio frequency (RF) components. At the bright port, the input laser is modulated at the radio frequency (RF) $f_{m_1}=\omega_{m_1}/2\pi$ . This RF sideband is resonant in the power-recycling cavity (PRC) and anti-resonant in the L-shaped resonator, serving as the local oscillator and carrier for sensing the common mode $L_+$ and PRC length, $l_p$, respectively. The common mode is sensed at the reflection port, whereas the PRC length, $l_p$, is sensed at the pick-off port. To minimize the coupling of two degrees of freedom at the two ports, hierarchy control is applied, similar to the strategy used in Advanced LIGO.  The differential mode, $L_-$, is sensed at the dark port utilising the so-called DC readout\,\cite{Ward_CQG2008,Fricke_2012}, where the local oscillator is introduced by placing a small offset of  $L_- + l_-$ .

In contrast to the strategy adopted by Advanced LIGO, our scheme directly introduces the local oscillator and carrier used to probe the $l_s$ from the dark port, without resorting to the introduction of Schnupp asymmetry\,\cite{f.Schnupp} to deliver the RF sideband into the signal-extraction cavity (SEC). Specifically, an auxiliary laser with a frequency shift of $\omega_{m_2}/2\pi$ relative to that of the main laser is employed at the dark port. It is phase-locked to the main laser and further modulated at  $\omega_{m_3}$ . In practice, one may also consider using the frequency doubling to create such an auxiliary laser, which may help prevent interference with the squeezed light injection that shares a similar optical path. The frequency of the auxiliary laser is resonant in the SEC and anti-resonant in the L-shaped resonator, ensuring that the light will only return to the dark port, as we will discuss in Sec.~\ref{sec:opt}. The $\omega_{m_3}$  sideband is anti-resonant in the SEC and arm cavity. In the output path, both components are filtered by an output mode cleaner (OMC)\,\cite{OMC_ligo2011}. At the OMC reflection port,  the demodulation at $\omega_{m_3}$ would produce a standard  \,\cite{PDH_1983}. Specifically, the modulation sideband of auxiliary laser contributes only to a DC component that couples with the AC spectrum generated by the carrier of auxiliary laser, which is modulated at $\omega_{m_2}$, making the readout solely dependent on the length of SEC. Moreover, the different resonant conditions between auxiliary laser and its sidebands ensure that the $l_s$ power spectrum is not cancelled.

Finally, a highlight point is that the control of {$L_- - l_-$} is not required. In Sec.~\ref{sec:opt}, we will present a qualitative analysis on the optical properties of the new configuration, elucidate why we remove Schnupp asymmetry and demonstrate the feasibility of leaving such differential degree of freedom uncontrolled .

\section{ Optical Properties}\label{sec:opt}
In this section, we study the optical properities of the interferometer with respect to carrier light and radio frequency sidebands required for the purpose of sensing and control.

\begin{figure}[htbp]
\centering

    \begin{minipage}[t]{0.492\linewidth}
        \centering
        \includegraphics[width=\textwidth]{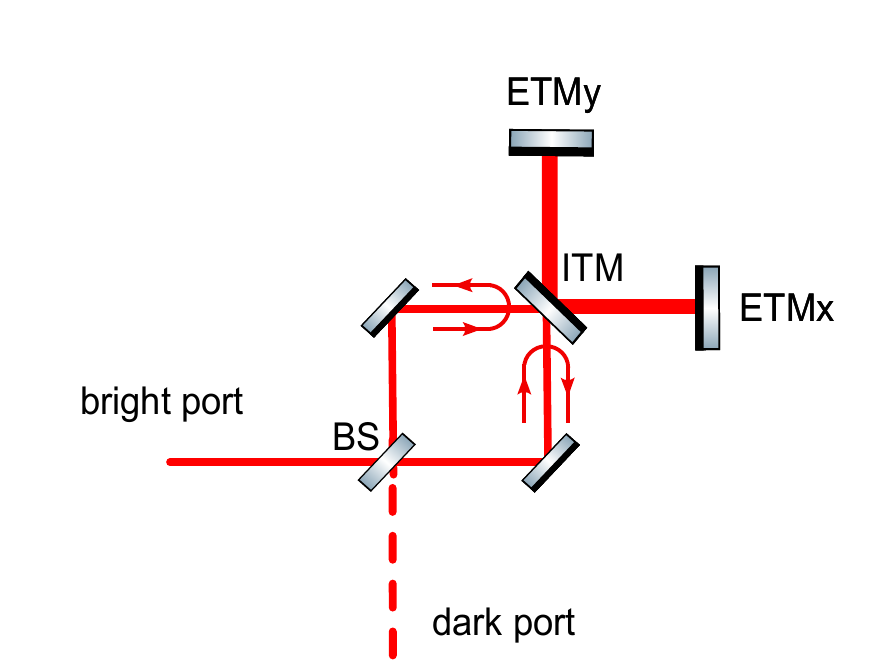}

    \end{minipage}
    \begin{minipage}[t]{0.492\linewidth}
        \centering
        \includegraphics[width=\textwidth]{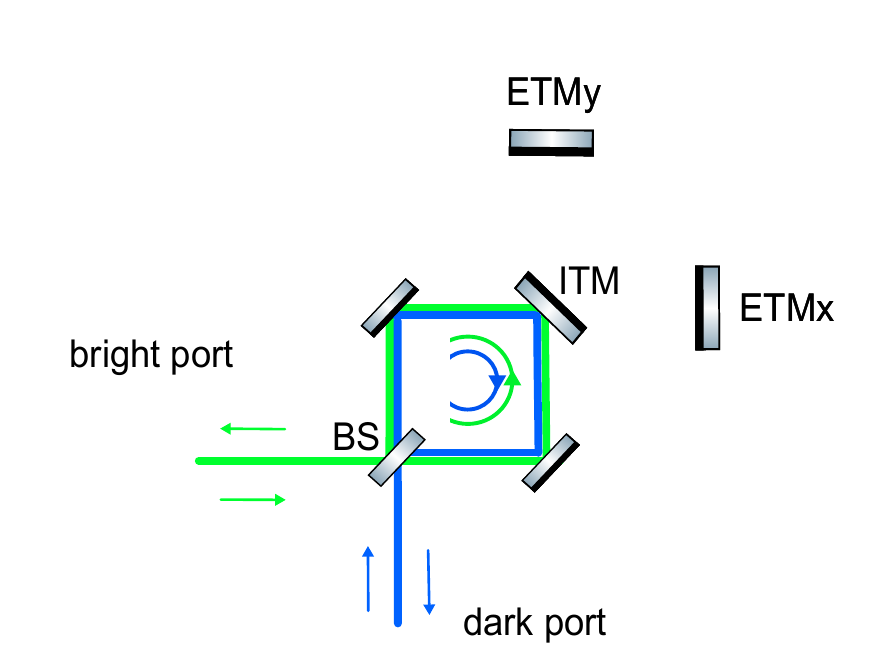}

    \end{minipage}
\caption{Schematic of two propagation modes in the new configuration. The left one shows the backward propagation mode when laser is injected from the bright side, which is similar to a Michelson interferometer. The right one shows the round-trip mode, which is identical to a Sagnac interferometer.}
\label{fig:modes}
\end{figure}

An investigation of the response of the system to the{ small time dependent sinusoidal fluctuation} of each longitudinal degree of freedom is highly necessary. Our analysis revealed that the common mode responses of the system at bright ports are indistinguishable from those of the Dual-Recycled Fabry-Perot Michelson Interferometer (DRFPMI) configuration. Meanwhile, it should be noted that two major differences exist between the new configuration and the traditional dual-recycled Fabry Perot Michelson interferometer. Firstly, the responses of the system to $L_-$ and $l_-$ are identical at low frequencies. Secondly, the interferometer exhibits Sagnac-like behavior for sidebands, which implies that the Schnupp asymmetry is not effective in the system.

the Optical properties of the new configuration involve both the L-shaped resonator and the central vortex. To gain a better understanding of its optical properties, {we conducted a calculation of quasi-static (DC) propagation of optical fields in the system}. The propagation of {quasi-static lights(DC lights)} can be classified into two modes: backward mode and round-trip mode. The backward mode describes light that enters from a path of the central vortex and travels back, whereas the round-trip mode refers to light that travels around the central loop, regardless of whether it enters the L-resonator. Fig.\,\ref{fig:modes} shows a schematic of the DC modes in the L-resonator and central vortex. These two propagation modes correspond to the two essential optical characteristics of the new configuration.

\subsection{Optical Characteristic of Carrier}
For the carrier, the L-resonator behaves like an impedance-matched cavity but with the front mirror and 
end mirror being the same ITM; the field reflected by the ITM and the field transmitted through the ITM cancel each other perfectly, leaving only the backward propagation field. Under the backward propagation mode, the central vortex and L-shaped resonator exhibit the characteristics of a Michelson interferometer with arm lengths $l_1+l_2+L_x$ and $l_3+l_4+L_y$, respectively.  
Under this propagation mode, the low frequency $L_+$ response is exactly the same as that of Fabry-Perot Michelson, and at low frequencies, the phase difference between the two effective Michelson arm of the backward mode is always proportional to $L_- + l_-$, indicating that the two differential degrees of freedom are degenerate at low frequencies.

\begin{figure}[t!]
\centering
\includegraphics[scale=0.7]{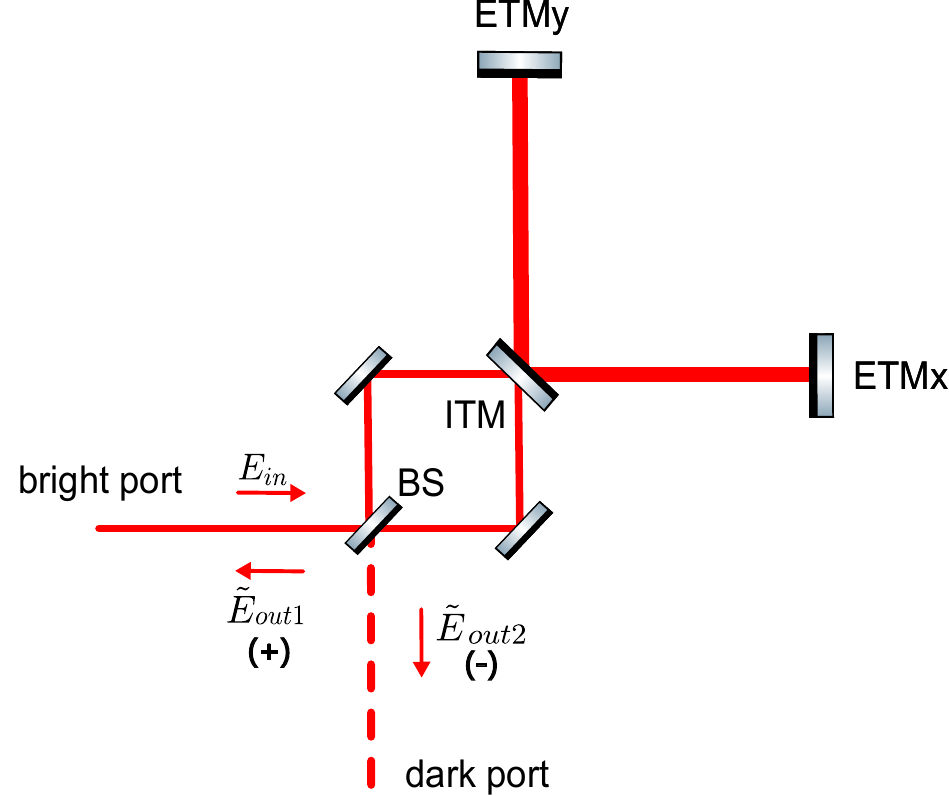}
\caption{A schematic of L-resonator together with central vortex. The common mode longitudinal signal can be obtained at the bright port, while the differential mode longitudinal signal can be obtained at the dark port. }
\label{fig:L-resonator}
\end{figure}

To explore the optical properties of the carrier more explicitly, we calculate the AC transfer functions of fields in the system. For simplicity, we assume that the system is balanced and all mirrors except the ITM are idealized.
The result shows that the transfer function for the common length degree of freedom $L_+$ is identical to that of Fabry-Perot Michelson Interferometer: 
\begin{equation}\label{eq12345}
{\cal T}(L_+)\equiv \frac{\partial{\tilde E_{\rm out1}}}{\partial{L_+}} =  {\frac{4\pi iE_{\rm in}}{\lambda}\frac{[r_{\rm i}(1+\mathrm{e}^{i\phi})+(1+r_{\rm i}^2\mathrm{e}^{i\phi})\mathrm{cos}(\delta \Psi)]\mathrm{e}^{\frac{i\phi}{2}}}{1-r^2_{\rm i}\mathrm{e}^{2i\phi}}}\,,
\end{equation}

where {$r_i$ is the amplitude reflectivity of ITM, $\phi=\frac{4\pi fL_+}{c}$ is the average phase delay of round-trip propagation in an arm of the L-shaped resonator, $\delta \Psi = \frac{2\pi f_0(L_-+l_-)}{c}$ is the phase difference of two propagation paths of light, here, $f$ is the frequency of signal, and $f_0$ denotes the frequency of laser.} Meanwhile, $E_{\rm in}$ is the input field, and $\tilde E_{\rm out1}$ is the reflected field at the bright port, as illustrated in Fig.\,\ref{fig:L-resonator}.

However, for the differential degrees of 
freedom $L_-$ and $l_-$, the response is 
drastically different from that of Michelson: 
\begin{equation}\label{eq123456}
{\cal T}(L_-) \equiv \frac{\partial{\tilde E_{\rm out2}}}{\partial{L_-}} =  {\frac{4\pi iE_{\rm in}}{\lambda}\frac{[r_{\rm i}(1-\mathrm{e}^{i\phi})+(1-r_{\rm i}^2\mathrm{e}^{i\phi})\mathrm{cos}(\delta \Psi)]\mathrm{e}^{\frac{i\phi}{2}}}{1-r^2_{\rm i}\mathrm{e}^{2i\phi}}}\,,
\end{equation}
\begin{equation}\label{eq1234567}
{\cal T}(l_-) \equiv \frac{\partial{\tilde E_{\rm out2}}}{\partial{l_-}} = {\frac{2\pi iE_{\rm in}}{\lambda}[r_{\rm i}+{\rm cos} \delta \Psi + \frac{t_{\rm i}^2\mathrm{e}^{i\phi}({\rm cos} \delta \Psi-r_{\rm i}\mathrm{e}^{i\phi})}{1-r^2_{\rm i}\mathrm{e}^{2i\phi}}}]\,.
\end{equation}
where {$t_{\rm i}$ is the transmissivity of ITM, $\tilde E_{\rm out2}$} denotes the reflected field at the dark port.

The transfer function of the L-resonator exhibits two notable features. Firstly, the resonant frequency of the longitudinal signal of the differential mode $L_-$ is located at $c/(4L_+)$, which deviates significantly from that of the common mode. This feature results in a high sensitivity around kHzs for the new configuration. {Secondly, The $L_-$ and $l_-$ degrees of freedom always appear together in the form of $L_- + l_-$, which actually depicts the disparity in optical path lengths between the two shortest routes in the backward propagation mode, while the difference of differential degrees of freedom $L_- - l_-$ is hidden in these results.}
Meanwhile, the $L_-$ and $l_-$ signals are indistinguishable at low frequencies when $L_- + l_-$ is tuned, as demonstrated by their ratio in Eq.\,\eqref{543}, which is consistent with the absence of $L_- - l_-$ in the low frequency behaviors in the system. 
\begin{equation}\label{543}
{\frac{{\cal T}(L_-)}{{\cal T}(l_-)} = \frac{2(1+r_{\rm i})\mathrm{e}^{i\phi/2}}{(1+r_{\rm i})(1+r_{\rm i}\mathrm{e}^{i\phi})+t_{\rm i}^2\mathrm{e}^{i\phi}}=\frac{1}{{\rm cos}(\phi/2)}\approx 1}\,.
\end{equation}

{The discussion above implies that $L_-$ and $l_-$ are not the best choice of degrees of freedom to describe the low-frequency optical behavior of the system, and the degeneracy between $L_-$ and $l_-$ indicates that we cannot lock them respectively. In fact, the sum of diffrential degrees of freedom, $L_- + l_-$ , and the difference between differential degrees of freedom, $L_- - l_-$ could capture the optical character of system more precisely. We propose that the control scheme should concentrate exclusively on actively regulating the composite degree $L_- + l_-$, while allowing $L_- - l_-$ to remain uncontrolled.}

{We would like to briefly demonstrate the feasibility of letting $L_- - l_-$ free swinging. Firstly, the discussion and calculation above indicates that the optical behaviors of carrier and error signals at low frequencies are insensitive to $L_- - l_-$, which means its fluctuation will not affect the feedback control of the rest degrees of freedom. }

\begin{figure}[t!]
\centering
\includegraphics[scale=0.5]{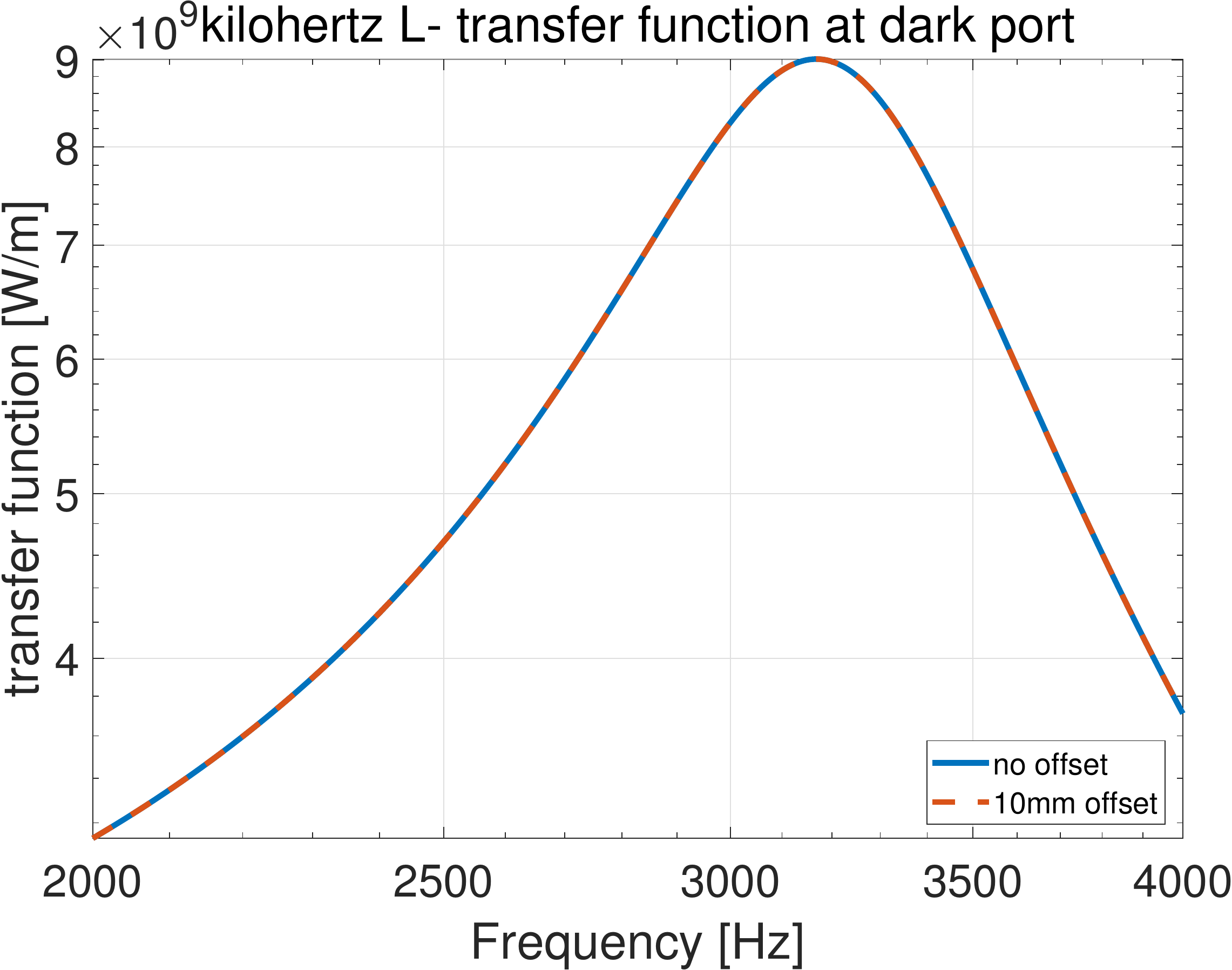}
\caption{Optical simulation result of kilohertz $L_-$ transfer function at dark port with $L_- - l_-$ offset. The result indicates that the gravitational wave readout is extremely insensitive to $L_- - l_-$ offset, making it feasible to leave it uncontrolled.}
\label{fig:offset}
\end{figure}

{Another concern is that  whether the DC offset or low frequency fluctuation affects the readout of GW signal at kilohertz. Intuitively, the L-shaped resonator actually flips the AC signal generating in the cavity by frequency $c/4L_x$ and $c/4L_y$, and the offset or fluctuation of $L_- - l_-$ would affect the final readout by moving the flipping frequencies. However, such effect becomes significant only when the $L_- - l_-$ offset is comparable to $L_+$, which is at the scale of kilometers. Based on this qualitative observation, we think that the final readout should be insensitive to low frequency fluctuation of $L_- - l_-$ either. We also conduct an Optickle simulation and find that the ultimate readout would remain the same even under millimeters of $L_--l_-$ offset, as shown in Fig.\,\ref{fig:offset}}

{In brief, although this system comprises five degrees of freedom, only the DC offset and low-frequency noise of four degrees of freedom, $L_+$, $l_p$, $l_s$, and $\cal {L}$ $= L_- + l_-$, significantly affect the system's optical behavior. Simultaneously, the low-frequency control of the system and the kHz-level signal readout are minimally sensitive to the low-frequency noise of $L_- - l_-$. This signifies that, in terms of system control, this degree of freedom effectively remains concealed, allowing it to be left uncontrolled without manifest side-effects. Consequently, in the design of active control schemes, we effectively control only the other four degrees of freedom. This simplification bears substantial importance for the overall control strategy. }

\subsection{Optical Properties of rf Sidebands}

Unlike the carrier, the modulation sidebands that are not resonant inside the arm cavity propagate in the round-trip mode, and the central vortex operates as a Sagnac interferometer. Specifically, when the RF sidebands are injected from the bright side, they cancel out at the dark port and do not enter the signal extraction cavity, regardless of whether the central vortex is balanced. Similarly, the sidebands introduced from the dark port cannot be transmitted to the bright port. Therefore, to acquire the sensing signal of the auxiliary length of the signal extraction cavity, introducing a non-zero Schnupp asymmetry is ineffective in this system. Consequently, injection from the dark side becomes the only viable option. This is the principal reason why we utilize dark port injection instead of the Schnupp technique in our sensing and control scheme. Consequently, the overall sensing control scheme is almost balanced, and the $l_{\rm s}$ signal is automatically disentangled from $L_+$ and $l_{\rm p}$, making it easier to diagonalize the sensing matrix.

Meanwhile, due to the Sagnac propagation mode of auxiliary fields, the low-frequency behaviors of these sidebands only envolve the common degrees of freedom, which has important guiding significance for the subsequent design of complete control scheme.

\section{Sensing matrix}\label{sec:transferfunctions}

In this section, we present the derivation of the sensing matrix, which establishes a connection between the demodulated signals at the readouts and longitudinal degrees of freedom. Given the control scheme illustrated in Fig.\,\ref{fig:config} and discussion above, we have 
\begin{equation*}
    \begin{gathered}
    \begin{pmatrix}\tilde P_{\rm refl}\\\tilde P_{\rm pop}\\\tilde P_{\rm omc}\\\tilde P_{\rm dc}\end{pmatrix}
    =
    \begin{pmatrix}\frac{\partial{\tilde P_{\rm ref}}}{\partial{L_+}}&\frac{\partial{\tilde P_{\rm ref}}}{\partial{l_p}}&0&0\\\frac{\partial{\tilde P_{\rm pop}}}{\partial{L_+}}&\frac{\partial{\tilde P_{\rm pop}}}{\partial{l_p}}&0&0\\0&0&\frac{\partial{\tilde P_{\rm omc}}}{\partial{l_s}}&0\\0&0&0&\frac{\partial{\tilde P_{\rm dc}}}{\partial{\cal L_-}}\end{pmatrix}
    \quad
    \begin{pmatrix}L_{+}\\l_{p}\\l_{s}\\\cal L_{-}\end{pmatrix}\,.
    \end{gathered}
    \end{equation*}
{where ${\cal L_-}=L_-+l_-$ is the sum of differential degrees of freedom.} The sensing matrix is nearly diagonal, primarily due to the absence of Schnupp asymmetry and the incorporation of modulation sidebands from both the bright and dark ports. The four power readouts are defined as follows:
\begin{enumerate}
\item $\tilde P_{\rm refl}$: In-phase readout with demodulation {of angular frequency $\omega_{m_1}$} at the reflection port, containing messages of $L_+$ and $l_p$ deviance.
\item $\tilde P_{\rm pop}$: In-phase readout with demodulation {of angular frequency $\omega_{m_1}$} at the pick-off port, containing messages of $L_+$ and $l_p$ deviance and linearly independent of $\tilde P_{\rm ref}$.
\item $\tilde P_{\rm omc}$: In-phase readout with demodulation {of angular frequency $\omega_{m_3}$} at the dark port, containing signal of $l_s$.
\item $\tilde P_{\rm dc}$: DC readout at the dark port, where the $L_-$ signal caused by gravitational waves can be read. {At low frequencies, readout at this port could also serve as $\cal L_-$ error signal.}
\end{enumerate}

In the following subsections, we will provide explicit expressions for the elements in the sensing matrix and compare them with numerical simulation with Optickle\,\cite{Optickle}. 

\subsection{Optical Preparation}

Preliminary work is required to calculate the transfer functions at different ports, some preliminary work is required. In this section, we provide a brief introduction to the audio-sideband perturbation technique \cite{Regehr_1995} that we employ for the computation of in-phase signals, and establish a set of variables to facilitate the subsequent calculations.

\subsubsection{Audio-sideband Perturbation}\par

 The audio-sideband perturbation technique is utilized to obtain the interferometric responses of an optical system as a function of frequency. This method considers the additional phase due to longitudinal fluctuations as a linear perturbation and expands the output electric field to the linear term in the Taylor series to obtain the transfer function of each degree of freedom at various frequencies. In this section, we apply this technique to deduce the principal formula of the transfer function under in-phase demodulation. The result will also serve as a guideline for the specific derivation of the bright port transfer functions in the following subsections. For simplification, we assume that the input laser has no laser noise. Hence, the input electric field after modulation can be represented as:
\begin{align}\label{eq134}
\tilde E_{\rm in} = E_0\mathrm{e}^{i(\omega t+\beta \sin\omega_{m}t)} = \mathrm{e}^{i\omega t}\left [E_{\rm c,in}+E_{\rm s,in}(\mathrm{e}^{i\omega_{m}t}-\mathrm{e}^{-i\omega_{m}t})\right]\,.
\end{align}
Here, $\omega$ is the angular frequency of the {carrier, 
$\omega_m$ is the angular frequency of RF modulator,and $\beta$ is the modulation depth.} Meanwhile, $E_{\rm c,in}$ and $E_{\rm s,in}$ are the input field components of the carrier and sideband fields, respectively. Without a loss of generality, we assume that both are real. By applying a linear approximation, the output electric field can be expressed as
\begin{align}\label{eq135}
\tilde E_{\rm out} = \mathrm{e}^{i\omega t}\left \{[\overline E_{\rm c}+\delta \tilde E_{\rm c}(t)]+[\overline E_{\rm s}+\delta \tilde E_{\rm s}(t)](\mathrm{e}^{i\omega_{m}t}-\mathrm{e}^{-i\omega_{m}t})\right\}\,,
\end{align}
where $\overline E_{\rm c}$ and $\overline E_{\rm s}$ are the DC field components of the carrier and sideband at the output respectively, $\tilde E_{\rm c}(t)$ and {$\tilde E_{\rm s}(t)$} are time-dependent field component caused by longitudinal fluctuation of the optical system. In Eq.\,\eqref{eq135}, we use the fact that the upper and lower sideband fields have exactly the same resonant condition in the system. Additionally, because the carrier and sideband are either resonant or anti-resonant in optical cavities, the DC field component of the carrier and sideband would be either a local maximum or local minimum, ensuring that both $E_{\rm c}$ and $E_{\rm s}$ are real. Moreover, {$\tilde E_{\rm c}(t)$} and $\tilde E_{s}(t)$ are pure imaginary numbers for tiny longitudinal fluctuations at low frequencies. The total laser power at the reflection port is simply $P_{out} =|\tilde E_{\rm out}|^2$, from which we obtain the explicit expression of power readout under in-phase demodulation $\sin(\omega_{m1}t)$:

\begin{align}\nonumber
\tilde P_{\rm out}(t) &=\frac{1}{2}  \textrm{Im} \left[\overline E_{\rm s}\tilde E^*_{\rm c}(t) +\overline E^*_{\rm c}\tilde E_{\rm s}(t)-\overline E^*_{\rm s} \tilde E_{\rm c}(t) -\overline E_{\rm c} \tilde E^*_{\rm s}(t)\right]\\ 
&\approx -i \left[\overline E^*_{\rm c}\tilde E_{\rm s}(t)-\overline E^*_{\rm s} \tilde E_{\rm c}(t)\right]\,,
\label{eq136}
\end{align}

By performing a Fourier transformation, we can derive the analytical expression for the in-phase power readout in the frequency domain
\begin{equation}\label{eq14}
\tilde P_{\rm out}(f) = -i\left[\overline E^*_{c}\tilde E_{s}(f)-\overline E^*_{s}\tilde E_{c}(f)\right]\,.
\end{equation}

From the analysis above, we can clearly see that the demodulated power signal is a direct consequence of the coupling between the DC field and the AC spectrum of the field, which is represented by two distinct terms. The first term originates from the coupling between the DC component of the carrier and the AC spectrum of the sideband, while the second term is a consequence of the AC coupling between the carrier and the DC component of the sideband. Consequently, we have succeeded in simplifying the problem by computing transfer functions to determine the DC and AC components of the output field.

\subsubsection{variable declaration}\par

For convenience, several parameters are introduced to characterize the behavior of the DC and AC fields.
For DC fields, the first set of parameters is the effective reflectivity of the L-shaped resonator for the carrier and sidebands.  
 In contrast to the previous section, optical loss in the arm cavity is taken into account by modeling the finite transmission, or non-unity reflectivity, of the end test masses (ETMs). Specifically, for the carrier, the effective reflectivity is given by 
\begin{align}\label{eq5}
r_{\textrm{L,\,$c_\pm$}}=-\frac{t^2_{\rm i}r_{\rm e}}{1\mp r_{\rm i}r_{\rm e}}\pm r_{\rm i}\,,
\end{align}
where $+$ and $-$ denote looking from the bright port and the dark port respectively , and $r_e$ is the {reflectivity} of the ETM.
For the sidebands, we have

\begin{align}\label{eq6}
r_\textrm{L,\,s$\pm$}=-\frac{t^2_ir_{\rm e} e^{i\phi_m}}{1\mp r_{\rm i}r_{\rm e}\mathrm{e}^{i\phi_m}}\pm r_{\rm i}\,,
\end{align}
where $\phi_m \equiv 4\pi f_m L_+/c$ denotes the modulation phase of the sidebands. For $+$ sign, $f_m \equiv f_{m1}$, while for $-$ sign, $f_m \equiv f_{m2}$. Another important set of quantities is the effective gain and reflectivity of the power recycling cavity (PRC) and signal extraction cavity (SEC). The expressions for the gains are as follows:
\begin{align}\label{eq9}
g_\textrm{prc,\,c}=\frac{t_{\rm p}}{1+t_{\rm po}^2r_{\rm p}r_\textrm{L,\,$c_+$}}, 
\; g_\textrm{prc,\,s}=\frac{r_{\rm p}}{1-t_{\rm po}^2r_{\rm p}r_\textrm{L,\,s+}},\; 
g_\textrm{sec,\,$c$}=\frac{t_{\rm s}}{1+r_{\rm s}r_\textrm{L,\,$c-$}},\; g_{\rm sec, s}=\frac{t_{\rm s}}{1+r_{\rm s}r_\textrm{L,\,s-}}\,, 
\end{align}
where $t_{\rm po}$ is the amplitude transmission coefficient of the pick-off mirror. The effective reflectivities are given by
\begin{align}\label{eq11}
r_\textrm{prc,\,c} &= r_{\rm p}+r_\textrm{L,$c_+$}t_pg_\textrm{prc,\,c} \;,r_\textrm{prc,\,s} = r_{\rm p}-r_\textrm{L,$s_+$}t_pg_\textrm{prc,\,s} \;,
r_\textrm{sec,\,s} = r_{\rm s}+r_\textrm{L,$s_-$}t_sg_\textrm{sec,\,s}\,.
\end{align}

Secondly,we introduce the parameters for describing the AC response of the system. The AC counterparts of the reflectivity of the L-resonator are  
\begin{align}\label{eq13}
\tilde r_\textrm{L,\,$c_{\pm}$}(f) &= -\frac{t^2_{\rm i}r_{\rm e}e^{i\phi}}{1\mp r_{\rm i}r_{\rm e}\mathrm{e}^{i\phi}}\pm r_i\,, 
\end{align}

\begin{align}\label{eq63d}
r_\textrm{L,\,s$\pm$}=-\frac{t^2_{\rm i}r_{\rm e} e^{i(\phi_m+\phi)}}{1\mp r_{\rm i}r_{\rm e}\mathrm{e}^{2i(\phi_m+\phi)}}\pm r_{\rm i}\,,
\end{align}
where $\phi\equiv 4\pi f L_+/c$ is the arm cavity phase of the audio-sideband signal at $f$. The PRC and SEC 
gains are 
\begin{align}\label{eq900}
\tilde g_\textrm{prc,\,$c_+$}=\frac{t_{\rm p}}{1+t_{\rm po}^2r_{\rm p}\tilde r_\textrm{L,\,$c$}}, 
\; \tilde g_\textrm{prc,\,s}=\frac{t_{\rm p}}{1-t_{\rm po}^2r_{\rm p}\tilde r_\textrm{L,\,s+}},\; 
\tilde g_\textrm{sec,\,$c$}=\frac{t_{\rm s}}{1+r_{\rm s}\tilde r_\textrm{L,\,$c-$}},\; \tilde g_{\rm sec, s}=\frac{t_{\rm s}}{1+r_{\rm s}\tilde r_\textrm{L,\,s-}}\,. 
\end{align}

With the aforementioned parameters, we shall deduce the analytical formulations for the $4 \times 4$ sensing matrix for both the common and differential degrees of freedom in the following subsections.

\subsection{Reflection Port}

Subsequently, our objective is to obtain the analytical expressions for these field components separately. We note that the DC field components of the carrier and sideband can be readily obtained and are simply given by the square root of the corresponding laser power. Simutaneously, the AC spectrum of the sideband field can be expressed in a straightforward manner, as the sideband has not entered the L-resonator and thus, its $L_+$ dependence is relatively weak. Accordingly, the AC spectrum of this sideband is proportional to the $l_p$ spectrum. In contrast, the AC spectrum of the carrier field is more intricate, as the carrier is also resonant in the L-resonator, therefore, it exhibits both $L_+$ dependence and $l_p$ dependence. The analytical expressions for the respective field components are presented as follows:
\begin{align}\label{eq15}
{\overline E_{\rm c,r}}=r_\textrm{prc,\,c} \sqrt{P_{\rm c}}
 \;,  {\overline E_{s,r}}=r_\textrm{prc,\,s}\sqrt{P_{\rm s}}\,,
\end{align}
\begin{align}\label{eq16}
\delta \tilde E_{\rm s,r}(f)=\frac{4\pi i }{\lambda}\sqrt{P_{\rm s}}g_{prc,\,s}g_{prc,\,s}(f)l_p(f)\,,
\end{align}
\begin{align}\label{eq17}
\delta \tilde E_{\rm c,r}(f)=\frac{4\pi i }{\lambda}
\sqrt{P_{\rm c}}g_\textrm{prc,\,c}g_\textrm{prc,\,c}(f)\left[l_p(f)+\frac{t^2_iM(f)}{2(1-r^2_ir^2_e)^2}L_+(f)\right]\,,
\end{align}
where $M(f)=\mathrm{e}^{i\phi}\left[1+r_ir_e+r_ir_e(1+r_ir_e)\mathrm{e}^{i\phi}\right]$ is the interferometric factor that depicts the common mode response of L-resonator.
By utilizing the expression for the power signal and employing the empirical relationship between the gain and bandwidth of the Michelson configuration, we can further simplify the theoretical expression for the power signal at the readout port
\begin{align}\nonumber
\tilde P_{\rm refl}(f)=\frac{4\pi }{\lambda}\sqrt{P_{\rm c}P_s}\left[g_\textrm{prc,\,c}^2r_\textrm{prc,\,s}\frac{t^2_iM(f)}{2(1-r^2_ir^2_e)^2}\right. &\frac{1}{1+s_{cc}}L_+(f)\\
\label{eq18}
&\left.+(g_\textrm{prc,\,c}^2r_\textrm{prc,\,s}-g_\textrm{prc,\,s}^2 r_\textrm{prc,\,c})\frac{1+s_r}{1+s_{cc}}l_p(f)\right]\,,
\end{align}
Here the chateristic frequencies are defined as 
\begin{equation}
    f_{\rm cc} = \frac{c}{16L_+}(1-r^2_ir^2_e)(1-r_pr_\textrm{L,$c_+$})\,,\quad s_{cc}=i\frac{f}{f_{cc}}\,,
\end{equation}
and 
\begin{equation}
    f_{\rm r}=f_{cc}\left(1-\frac{g_\textrm{prc,\,c}^2r_\textrm{prc,\,s}}{g_\textrm{prc,\,s}^2r_\textrm{prc,\,c}}\right)
    \,,\quad s_{\rm r}=i\frac{f}{f_r}\,.
\end{equation}

The sensitivity of the length sensing control is influenced by the transmissivity of the power recycling mirror, denoted as $t^2_p$. In our design, we set $t^2_p$ as 0.014 to ensure that the power recycling cavity is critically coupled for the carrier. This choice of parameter has several advantages. Firstly, it allows the reflection coefficient of the power recycling mirror to be adjusted to achieve critical coupling for the carrier, which can decrease the overall DC reflection power and reduce the shot noise at the reflection port \cite{Izumi_CQG2017}. Additionally, the critically coupled power recycling mirror provides another advantage whereby the power spectrum at the readout port is primarily dominated by the $L_+$ term under low frequency. This effect can be attributed in part to the reflection behavior of the L-resonator, where the effective reflectivity for light that does not enter the arm cavities is much closer to unity than for those resonant inside the cavity. Consequently, the gain factor of the sideband, $g_{sb,1}$, is larger than that of the carrier, $g_{cr,1}$, which implies that the $l_p$ signal is more prominent in the AC spectrum of the RF sideband. By coupling the AC part of the RF sideband with a zero DC reflection field of the carrier, the $l_p$ sensitivity at this port can be partly suppressed, making the $L_+$ signal the dominant term. This approach enhances the diagonalization level of the final sensing matrix.

\begin{figure}[htbp]
\centering

    \begin{minipage}[t]{0.7\linewidth}
        \centering
        \includegraphics[width=\textwidth]{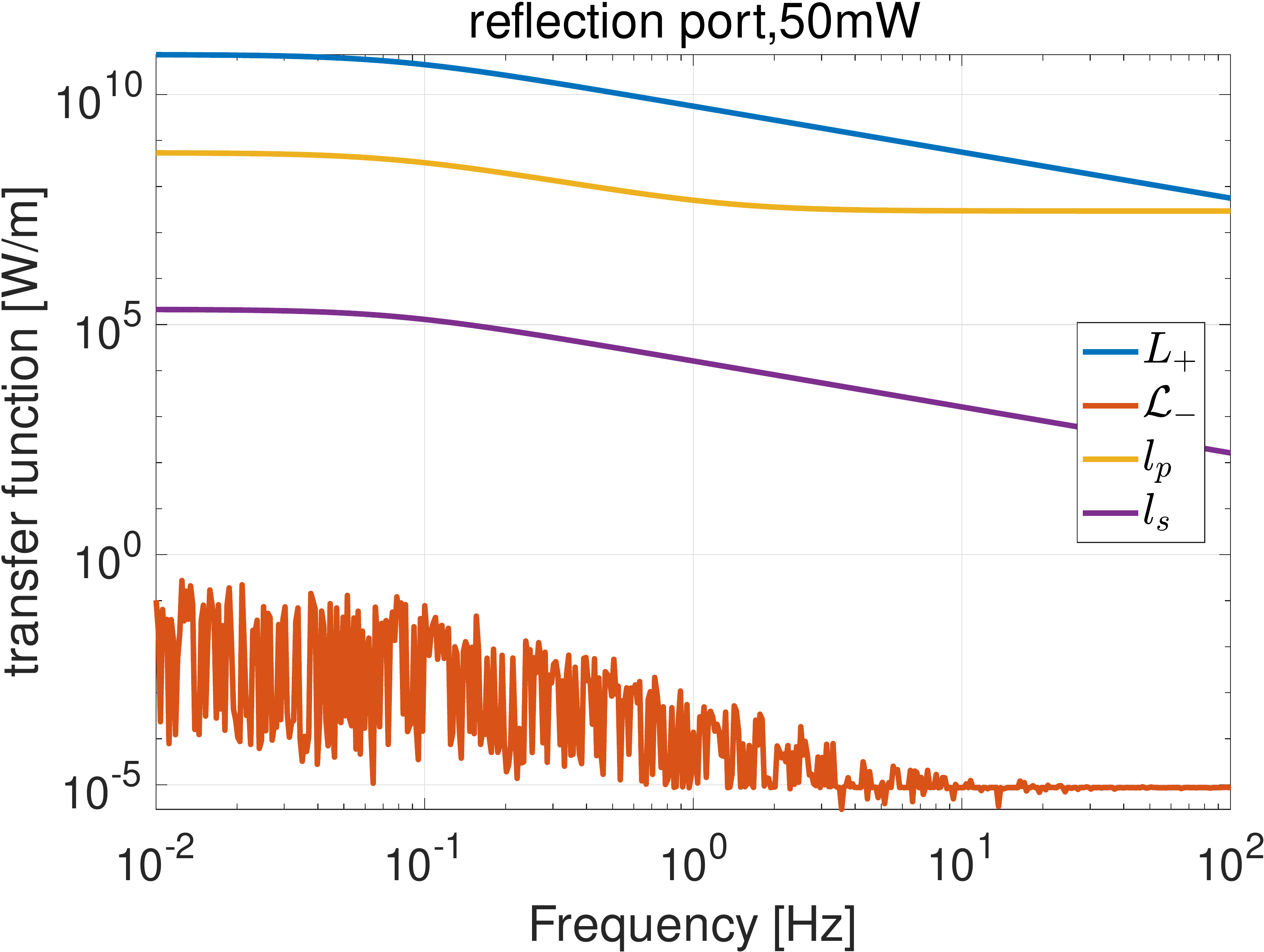}

    \end{minipage}

\caption{Transfer functions at the reflection port,an attenuator is applied to reduce the total power of the corresponding port to about 50mW. The picture is Optickle simulation result of transfer functions at the reflection port, from which we could conclude that $L_+$ signal is the dominant term under low frequency. Simulation result is in good correspondence with theoretical value, the difference is less than 1 percent.}
\label{fig:refl}
\end{figure}

In Fig.\,\ref{fig:refl},  the theoretical transfer functions at the readout port are plotted using MATLAB, and the resulting curve exhibits good correspondence with the simulated values, particularly under low frequency conditions.\par

\subsection{Pick-off Port}\par
By applying the same technique, we derive transfer functions at pick-off port analytically. The relative field components have the following form:

\begin{align}\label{eq19}
{\overline E_{\rm c,p}} = r_\textrm{po}g_\textrm{prc,\,c}\sqrt{P_{\rm c}} \;,
{\overline E_{\rm s,p}} = r_\textrm{po}g_\textrm{prc,\,s}\sqrt{P_{\rm s}} \,,
\end{align}

\begin{align}\label{eq20}
\delta \tilde E_{\rm s,p}(f) &= \frac{r_\textrm{po}}{t_p}\frac{4\pi i}{\lambda}\sqrt{P_{\rm s}}g_\textrm{prc,\,s}g_\textrm{prc,\,s}(f)l_{p}(f) \,,
\end{align}

\begin{align}\label{eq21}
\delta \tilde E_{\rm c,p}(f) &= \frac{4\pi i r_\textrm{po}}{\lambda t_p}g_\textrm{prc,\,c}g_\textrm{prc,\,c}(f)\sqrt{P_{\rm c}}\left[l_p(f)+\frac{t^2_iM(f)}{2(1-r^2_ir^2_e)^2}L_+(f)\right]\,.
\end{align}

We introduce the amplitude reflection rate of the pick-off mirror denoted as $r_\textrm{po}$. Utilizing the analytical expressions for the field components, we derive an analytic expression for the power readout. The demodulated power readout at the readout port is given by

\begin{align}\nonumber
\tilde P_{\rm pop}(f) = \frac{4\pi r^2_\textrm{po}}{\lambda t_p}\sqrt{P_{\rm c}P_{\rm s}}\left[g_\textrm{prc,\,c}^2g_\textrm{prc,\,s}\frac{t^2_iM(f)}{2(1-r^2_ir^2_e)^2}\frac{1}{1+s_{cc}}\right. &L_+(f)\\
\label{eq22}
+g_\textrm{prc,\,c}g_\textrm{prc,\,s}& \left.(g_\textrm{prc,\,c}-g_\textrm{prc,\,s})\frac{1+s_p}{1+s_{cc}}l_p(f)\right]\,.
\end{align}

Here the newly introduced characteristic frequency is defined as 
\begin{equation}
   f_{\rm p}=(1-\frac{g_\textrm{prc,c}}{g_\textrm{prc,s}})f_{cc} \,,\quad s_p=\frac{if}{f_p}\,,
\end{equation}

\begin{figure}[htbp]
\centering

    \begin{minipage}[t]{0.7\linewidth}
        \centering
        \includegraphics[width=\textwidth]{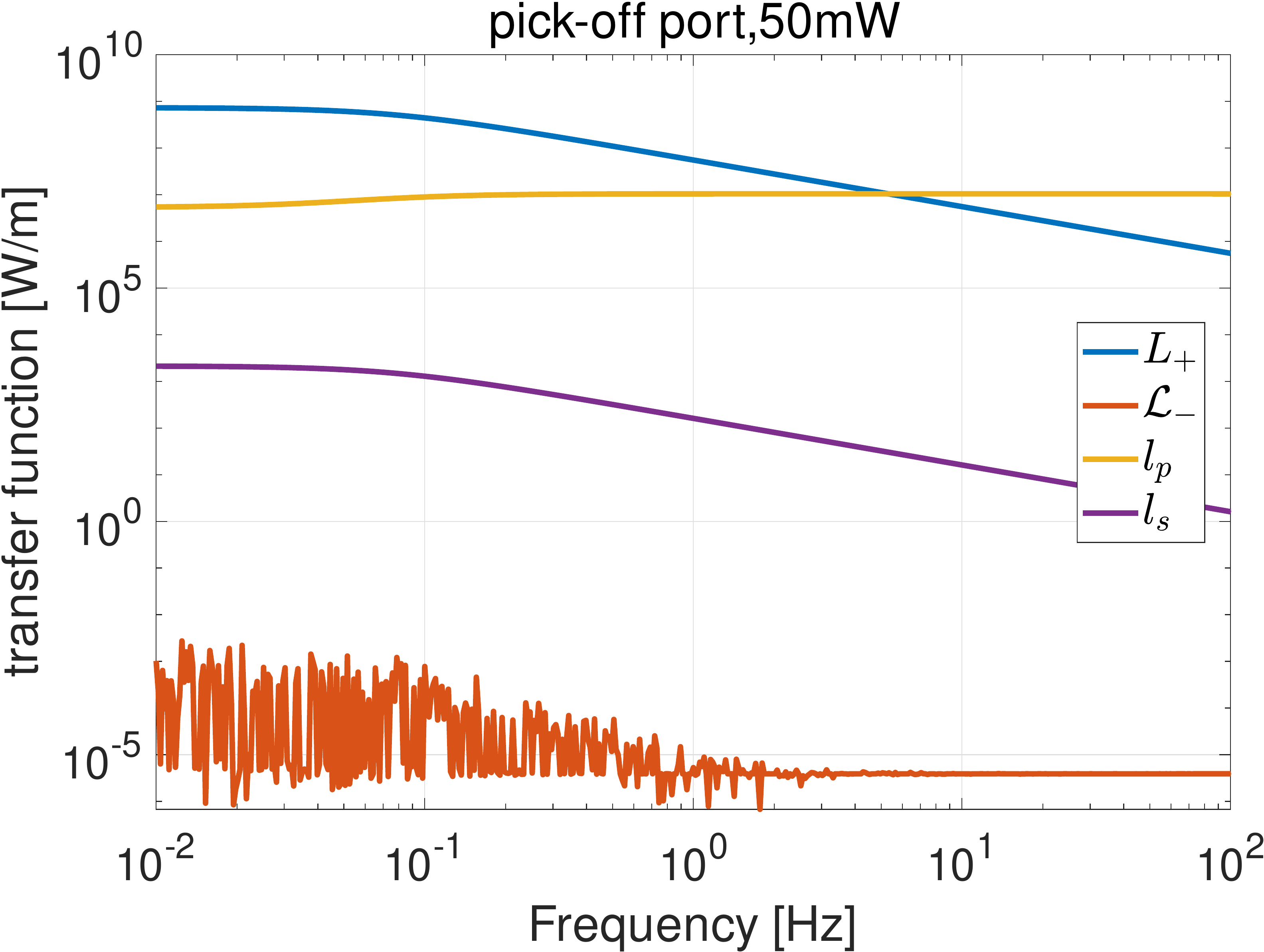}

    \end{minipage}

\caption{Transfer functions at the pick-off port. This figure is Optickle simulation result of transfer functions at the pick-off port, from which we could conclude that readout there is dominated by the $L_+$ and $l_p$ signal. Simulation result is in good correspondence with theoretical value, the difference is less than 1 percent.}
\label{fig:pick-off}
\end{figure}

In Fig.\,\ref{fig:pick-off}, the theoretical transfer functions are plotted, and the difference between the theoretical value and simulation result was less than 1 percent. Unlike the readout at the reflection port, the $l_{p}$ transfer function is comparable to the $L_+$ transfer function at this port even at low frequencies and surpasses the $L_+$ transfer function at approximately 3 Hz. Therefore, it can be concluded that the $l_{p}$ signal is rich in this port.

\subsection{Linear Combination}\par
Through a linear combination of readouts of the reflection and pick-off ports, we can obtain readout signal that is only proportional to longitudinal deviance of $l_p$ degree of freedom:
\begin{align}\label{eq23}
\tilde P_{\rm pop}(f)-\frac{g_\textrm{prc,\,s}r^2_\textrm{po}}{r_\textrm{prc,\,s}t_p}\tilde P_{\rm refl}(f)=\frac{r^2_\textrm{po}}{t_pr_\textrm{prc,\,s}}\frac{4\pi }{\lambda}\sqrt{P_{\rm c}P_{\rm s}}g_\textrm{sb1}^2(g_\textrm{prc,\,c}r_\textrm{prc,\,s}-g_\textrm{prc,\,s}r_\textrm{prc,\,c})l_p(f)\,.
\end{align}

Noticing that the $l_p$ transfer function after the linear combination has no frequency dependence, which might simplify the sensing control of the system. A method for decoupling $L_+$ signal and $l_p$ signals has already been developed\,\cite{Fritschel_Appl2001}. With the readout at the reflection port maintained as 0, the readout at pick-off port would be proportional to the deviance of single $l_p$ deviance.\par 

\subsection{OMC Reflection Port}\par
Denoting the power of auxiliary laser as $P_a$ , and defining a new parameter $P_{s_2}$ to refer to the power of the new sideband, then taking advantage of the fact that the effective reflection coefficient of the L-resonator for this sideband is very close to 1, we cab derive the theoretical expression for this non-zero $l_s$ transfer function as follows:

\begin{align}\label{eq31}
\frac{\partial \tilde P_{\rm omc}}{\partial l_s}=\frac{8\pi}{\lambda}r_{\rm s}g_\textrm{sec,\,s}^2\sqrt{P_{\rm a}P_{\rm s_2}}\,.
\end{align}

\begin{figure}[htbp]
\centering

    \begin{minipage}[t]{0.7\linewidth}
        \centering
        \includegraphics[width=\textwidth]{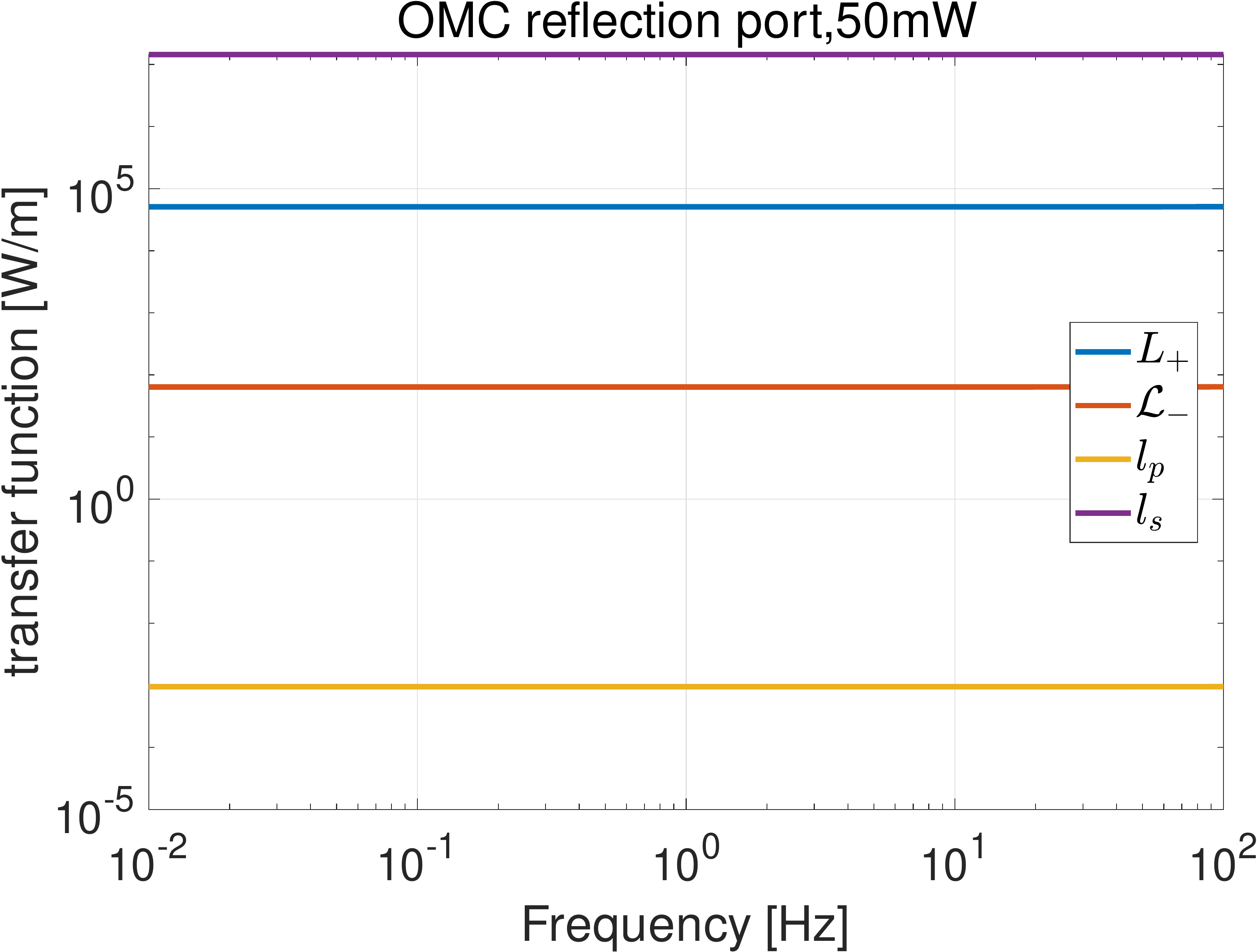}
        \centerline{}
    \end{minipage}

\caption{Transfer functions at the OMC reflection port, an attenuator is applied to reduce the total power of the corresponding port to about 50mW. The picture is Optickle simulation result of transfer functions at the in-phase readout at OMC reflection port, under demodulation frequency of beating frequencies of two sidebands. At this port, we could get a purified $l_s$ signal. Simulation result is in good correspondence with theoretical value, the difference is less than 1 percent.}
\label{fig:OMC2}
\end{figure}

Fig.\,\ref{fig:OMC2} displays the Optickle simulation results of the in-phase transfer functions at this readout port, from which we can conclude that the new $l_s$ transfer function significantly surpasses that of other transfer functions at the same port. 
The results show that we can obtain a pure $l_s$ signal with a transfer function up to $10^7$ W/m without any pre-operation or hierarchical control.

\subsection{Dark Port, DC}
At the dark port, power readout is dominated by the differential degrees of freedom. For gravitational wave detection, it is necessary to preserve the wide-range frequency dependence of the transfer functions at this port. Only two non-zero transfer functions, namely $L_-$ transfer function and $l_-$ transfer function, are present at this readout port. Specifically, the partial derivative of $P_4(f)$ with respect to $L_-$ and $l_-$ are given by 

\begin{align}\label{eq29}
\frac{\partial{\tilde P_{\rm dc}(f)}}{\partial{L_-}}=\frac{4\pi }{\lambda}P_{\rm c}sin{\phi_-}g_\textrm{prc,\,c}^2g_\textrm{sec,\,c}g_\textrm{sec,\,c}(f)\frac{t^2_iN(f)}{(1-r^2_ir^2_e)(1-r^2_ir^2_e\mathrm{e}^{2i\phi})}\,,
\end{align}
 
\begin{align}\label{eq30}
\frac{\partial{\tilde P_{\rm dc}(f)}}{\partial{l_-}}=-\frac{4\pi}{\lambda}P_{\rm c}\sin{\phi_-}g_\textrm{prc,\,c}^2g_\textrm{sec,\,c}g_\textrm{sec,\,c}(f)\tilde r_\textrm{L,\,$c_{-}$}(f)\,,
\end{align}

where $N(f)=\mathrm{e}^{i\phi}\left[1+r_ir_e-r_ir_e(1+r_ir_e)\mathrm{e}^{i\phi}\right]$ is the interferometric factor depicting the differential mode response of L-resonator. As we point out in Sec.\,\ref{sec:opt}, they're degenerate under low frequency, and vary significantly around kHzs. Notably, at low frequency, the DC readout exhibits almost no reliance on the common degree of freedoms due to a phase difference of $\frac{\pi}{2}$ with the transmitting carrier DC, which ensures that they do not contribute to the total DC power.

{The readout at dark port is actually a mixture of $L_-$ and $l_-$ signals, but there is no need to separate them. Around $f = \frac{c}{4L_+}$, the $L_-$ detecting sensitivity at dark port is much higher, making the $L_-$ signal of gravitational wave takes the absolute dominance. Meanwhile, at low frequencies, $L_-$ and $l_-$ are indistinguishable, and the readout could serve as an error signal of ${\cal L_-}=L_-+l_-$. The relative transfer function could be approximately expressed as:
\begin{align}\label{eq300}
\frac{\partial{\tilde P_{\rm dc}(f)}}{\partial{\cal{L_-}}}=-\frac{4\pi}{\lambda}P_{\rm c}g_\textrm{prc,\,c}^2g_\textrm{sec,\,c}^2 r_\textrm{L,\,$c_{-}$}\sin{\phi_-}\,,
\end{align}}

\begin{figure}[htbp]
\centering

    \begin{minipage}[t]{0.7\linewidth}
        \centering
        \includegraphics[width=\textwidth]{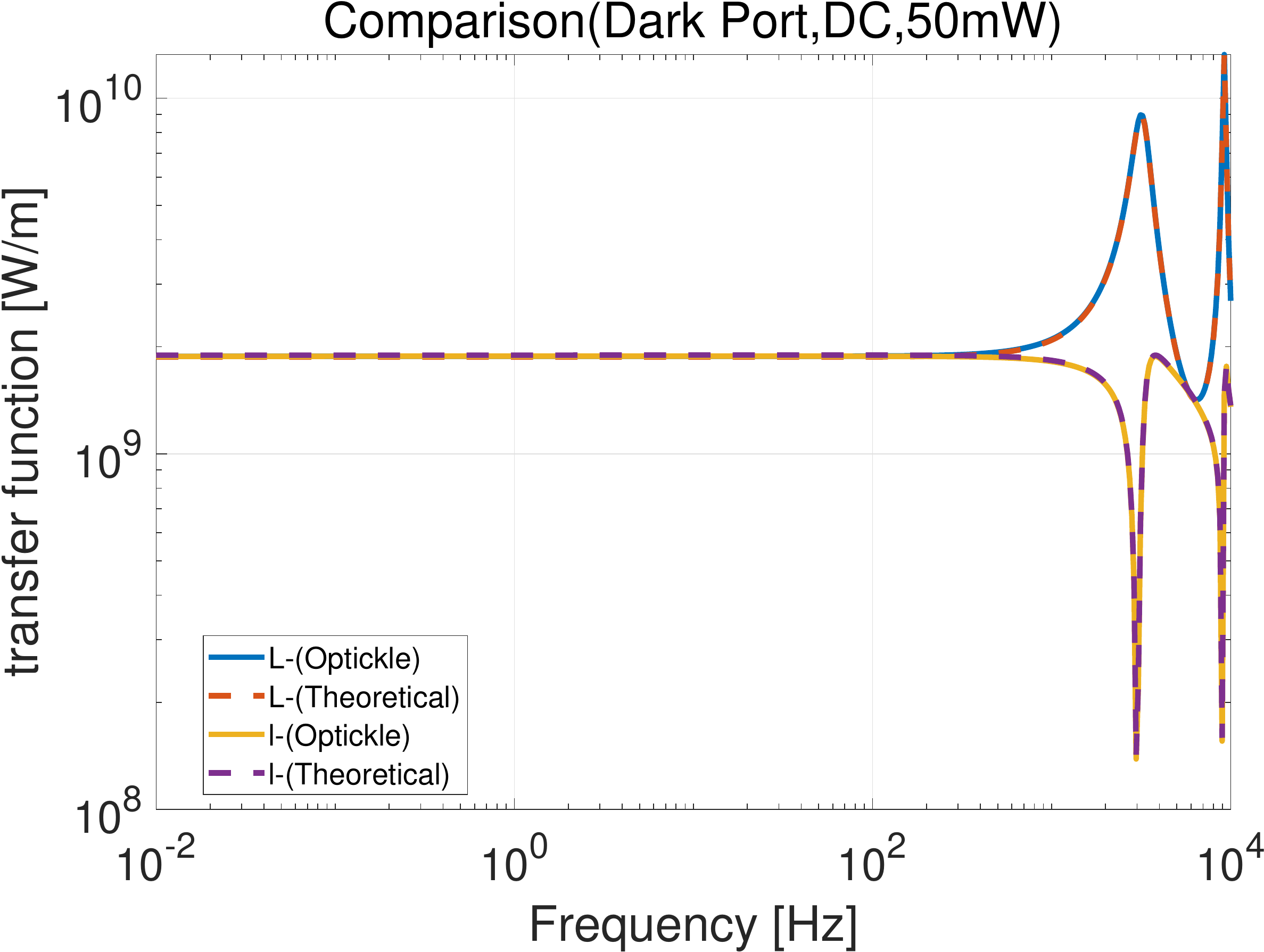}
    \end{minipage}

\caption{$L_-$ and $l_-$ transfer functions at the dark port. The theoretical and simulation results are in good agreement over a wide range, and at low frequencies, the difference is about 1 percent.}
\label{fig:dark}
\end{figure}

We utilize Optickle to simulate the transfer functions at this readout port, and present the simulation results of the transfer functions together with their theoretical values in Fig\,\ref{fig:dark}. While some differences between the two results are observed, the overall behavior remains consistent. From the picture, we could see that the SEC enhances the effective bandwidth of the gravitational wave (GW) detector around 3 kHz, where the light is anti-resonant in the L-resonator. This broadens the effective detection bandwidth to more than 100 Hz. The interferometric factor of the $L_-$ signal, denoted as $N(f)$, ensures that the system maintains sufficient sensitivity in this frequency range. Degeneracy between two differential degrees of freedom can also be observed in the figure, which is consistent to our previous discussion. At the low-frequency limit, the transfer functions of $L_-$ and $l_-$ remain nearly constant, which is advantageous for improving the sensing control quality.

\subsection{Numerical Sensing Matrix and Diagonalization Procedure}\label{sec:sensingmatrix}

 The sensing matrix reveals that the sensing signals obtained from the dark ports are automatically diagonalized, 
 while the $2\times 2$ block to lock the $L_+$ and $l_p$ degrees of freedom has two off-diagonal elements. When the PRC is critically coupled for the carrier of the main laser, the off-diagonal element in the first row is relatively small compared with the diagonal $L_+$ element, resulting in a nearly lower-diagonal block. With the analytical result of all the non-zero transfer functions, the entire sensing matrix {at low frequencies( $<$ 100Hz)} can be expressed numerically as follows(in the unit of W/m):

\begin{equation*}
    \begin{gathered}
A=\begin{pmatrix}7.3\times 10^{10}\times \frac{1}{1+if/0.047}&5.4\times 10^8\times \frac{1+if/1.30}{1+if/0.047}&\approx 0&0\\7.3\times 10^8\times\frac{1}{1+if/0.047}&5.4\times 10^6\times\frac{1+if/0.023}{1+if/0.047}&0&0\\0&0&1.5\times 10^7&0\\0&0&0&1.9\times 10^9\end{pmatrix}
 \end{gathered}
\end{equation*}

To obtain a diagonalized sensing matrix, a linear combination of the two readouts at the bright ports is necessary. Technically, this can be achieved by getting the signal at the pick-off port with the in-phase readout at the reflection port maintained as zero in advance\,\cite{Fritschel_Appl2001}. The numerical sensing matrix after a single diagonalization procedure is as follows:

\begin{equation*}
    \begin{gathered}
A=\begin{pmatrix}7.3\times 10^{10}\times \frac{1}{1+if/0.047}&5.4\times 10^8\times \frac{1+if/1.30}{1+if/0.047}&\approx 0&0\\0&1.0\times 10^7&0&0\\0&0&1.5\times 10^7&0\\0&0&0&1.9\times 10^9\end{pmatrix}
 \end{gathered}
\end{equation*}

At low frequencies (below 1Hz), readout at the reflection port is mainly governed by the $L_+$ transfer function, rendering the only off-diagonal element insignificant. Consequently, the sensing matrix becomes diagonal, enabling us to sense and control the four degrees of freedom separately.



\section{Conclusion and outlook}\label{sec:conclusion}

We have presented a preliminary study of the sensing
and control scheme for the new interferometer 
configuration based upon an L-shaped resonator. Such an inteferomter has a unique amplification effect on signals around kHz, which leads to higher sensitivity compared to the Michelson configuration for detecting the merger signals from binary neutron stars. Based on the optical characteristic of the new system, that is, degeneracy between two differential degrees of freedom under low frequency and Sagnac-like behavior for modulation sidebands , we design a sensing control scheme with one less degrees of freedom to control. We remain the sensing and control method of $L_+$ and $l_p$ degrees of freedom similar to that in Advanced LIGO, but remove the Schnupp asymmetry and use dark port injections to get the $l_s$ sensing signal instead. The gravitational-wave signal could be obtained at the dark port through a DC readout technique. 
Analytical derivation of the sensing matrix and comparison with numerical simulations  are also presented, showing that the sensing matrix of the new configuration is more easily diagonalizable and has a minimum non-zero transfer function of the order of $10^7$ W/m. 

This study serves as a starting point for future discussions on this new type of gravitational wave detector. There are still many issues remained to be considered. So far, we have not considered the radiation pressure effect on the dynamics of the test masses\,\cite{Somiya_PRD2006,Sheard_PRA2004,Evans_PRL2015}, and the laser noise\,\cite{Craig_arxiv2021} has also not been included. These are interesting and important topics for future research. 

\section{Acknowledgements}
We thank Chunnong Zhao, Matthew Evans, and AIC group for fruitful
discussions. X. G. and H. M. are supported by State Key Laboratory
of Low Dimensional Quantum Physics and the start-up
fund from Tsinghua University.  T. Z., D. M. acknowledge
the support of the Institute for Gravitational Wave Astronomy at the University of Birmingham, STFC Quantum Technology for Fundamental Physics scheme (Grant
No. ST/T006609/1), and EPSRC New Horizon Scheme
(Grant No. EP/V048872/1 ). T. Z. acknowledges the
support of department of Gravitational Waves and Fun-
damental Physics in Maastricht University and ETEST
project. D. M. is supported by
the 2021 Philip Leverhulme Prize.
\newpage
\section*{Appendix A: Parameters }\par

The table shown below summarizes the relevant optical parameters used in this paper. 
\begin{center}
\begin{tabular}{cccc}
\hline
Symbol& Description& Value& Units\\
\hline
$t^2_{\rm i}$& Power transmission of input mirror& 0.014 &\\
$t^2_{\rm e}$& Power transmission of end mirror& 0.00008 &\\
$t^2_{\rm p}$& Power transmission of power recycling mirror& 0.021 &\\
$t^2_{\rm s}$& Power transmission of signal extraction mirror& 0.06 &\\
$r_{\rm po}$& AR reflection rate of the pick-off port &$10^{-5}$& \\
\hline
Distance\\
$L_x,y$& Arm cavity length& 25000& m\\
$l_p$&Power recycling cavity length& 50.00& m\\
$l_s$& Signal extraction cavity length& 50.00& m\\
$l_{offset}$&  $l_-$ offset& 0.1064& nm\\
\hline
Laser properties \\
$P_{\rm in}$& Input laser power& 250& W\\
$P_{\rm a}$& DC power of the auxiliary laser &25 &mW\\
$P_{\rm c}$& Input power of carrier &248 &W \\
$P_{\rm s}$& Input power of the 1st rf sideband &1.2 &W \\
$P_{\rm s_2}$ & Input power of the 2nd rf sideband&12.5 &mW\\
$\Gamma$& Modulation depth of the 1st rf sideband& 0.1 & \\
$\omega_{m_1}/2\pi$& Modulation frequency of the 1st rf sideband& 13.5 &MHz\\
$\omega_{m_2}/2\pi$& Frequency shift of auxiliary laser& 39.0 &MHz\\
$\omega_{m_3}/2\pi$& Modulation frequency of the 2nd rf sideband & 22.5 &MHz\\
$\lambda$& Laser wavelength& 1064& nm\\
\hline
Laser power inside cavities \\
$P_{\rm p}$& Power inside power recycling cavity& 11 & kW\\
$P_{\rm L}$& Power inside the L-resonator& 1.5 & MW\\
$P_{\rm dark}$&Carrier at the dark port& 140 & mW\\
\hline
\end{tabular}
\end{center}
\par
\par

\section*{Appendix B: Overview of the sensing and control scheme of Advanced LIGO}\par

\begin{figure}[ht]
\centering
\includegraphics[scale=0.6]{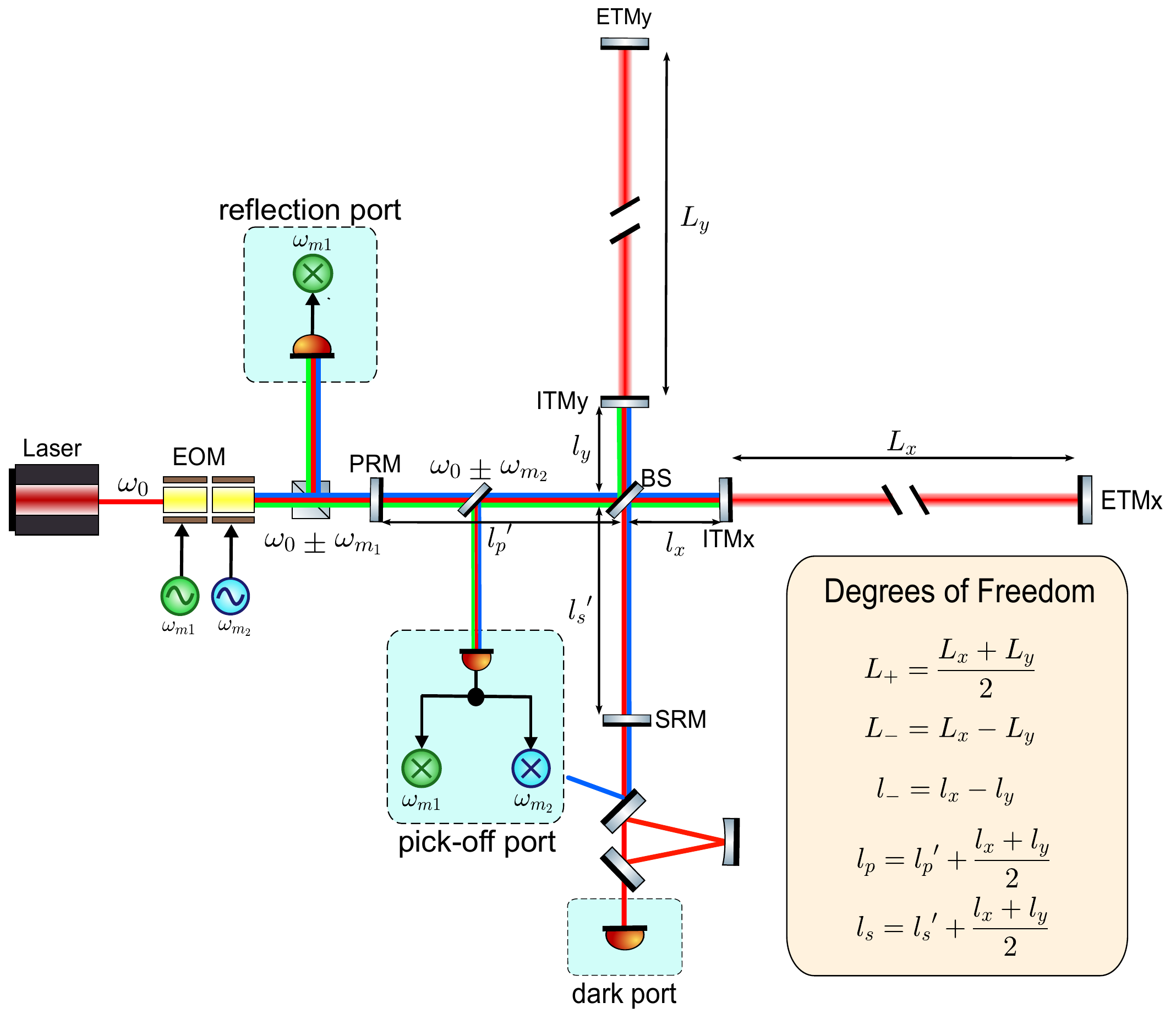}
\caption{A simplified optical layout of the Advanced LIGO and its sensing and control scheme\,\cite{Izumi_CQG2017}}.
\label{fig:Mich}
\end{figure}

In this appendix, we provide an overview of the Advanced LIGO sensing and control scheme, as presented in Ref.,\cite{Izumi_CQG2017}. This reference also establishes the design principles for the control scheme of the new configuration. Figure \ref{fig:Mich} illustrates the optical layout of the Advanced LIGO, which consists of five longitudinal degrees of freedom: $L_+$ and $L_-$, representing the average length and length difference of the arm cavities, $l_p$ and $l_s$, indicating the effective length of the power recycling cavity and signal recycling cavity, and $l_-$, denoting the length difference of the two Michelson arms. To maintain the integrity of the Michelson configuration, it is sensible to avoid introducing sensing ports in the Michelson arms or arm cavities. As a result, there are only three potential locations for positioning the photo-detector: the reflection port, the pick-off port, and the dark port.

The signals of the common degrees of freedom and the differential degrees of freedom are naturally separated from each other. Ideally, we would like to read out the signals of $L_+$ and $l_p$ at the bright port, while the signals of $L_-$, $l_-$, and $l_s$ would be read out at the dark port to maximize the sensitivity of length control. However, if we read the sensing signal of both differential degrees of freedom at the dark port, it becomes challenging to decouple these two degrees of freedom. To address this issue, we have adopted the technique of Schnupp asymmetry. By introducing a non-zero Schnupp asymmetry, the bright port becomes a preferable location for detecting the $l_-$ signal because there is no $L_-$ signal present, and the $l_-$ signal is under a quadrature phase. Furthermore, the information of the common degrees of freedom is encoded in the power signals through in-phase demodulation. Hence, the power readout at bright ports, under quadrature demodulation, relies solely on the single $l_-$ degree of freedom, resulting in a high level of diagonalization in the sensing matrix. To maximize the sensing signal of $l_-$, the sideband should be resonant in the power recycling cavity (PRC) and signal recycling cavity (SRC), while being anti-resonant in the arm cavities. Additionally, the DC readout technique assists in obtaining an almost purified $L_-$ signal.

Schnupp asymmetry provides another option for reading the $l_s$ signal. We can acquire the $l_s$ signal either at the dark port or through the in-phase readout at the bright port. Theoretical calculations suggest that the latter option offers higher sensitivity. Thus, the sensing signals of the three common degrees of freedom are encoded in the in-phase power readouts at the reflection port and the pick-off port. To obtain three independent in-phase power readouts at these ports, at least two RF sidebands are required. Consequently, an additional control sideband is necessary to fully lock the system. For optimal independence, this new sideband should be resonant in the PRC and anti-resonant in the SRC.

In summary, three in-phase power readouts at the bright ports, including one at the reflection port and two readouts at the pick-off port, are responsible for sensing the three degrees of freedom, $L_+$, $l_p$ and $l_s$. Meanwhile, the $l_-$ signal is sensed through the quadrature power readout at the pick-off port, and GW signal $L_-$ is obtained at the dark port through the DC readout.

\section*{Appendix C: Detailed derivation of transfer functions of L-shaped resonator}\par
{This is a new appendix to help the readers to follow results in Sec 3}

\setcounter{equation}{0}
\renewcommand{\theequation}{C.\arabic{equation}}
In this appendix, we present a comprehensive derivation of the transfer functions ($L_+$ and $L_-$) for the fields within the L-shaped resonator, as expressed in equations (6) and (7). This detailed derivation serves as a valuable resource for readers seeking a deeper comprehension of the system.

Our calculations are grounded in a fundamental principle: when a monochromatic light wave, represented as $E_0e^{i2\pi ft}$, impinges upon a mirror with a reflectivity of $r$ and undergoes a minute displacement $x(t)$ before being reflected, it perturbs the position of the reflection interface. Consequently, a time-dependent phase shift is induced in the reflected field, leading to an AC field spectrum within the reflected component:

\begin{equation}\label{C_1}
\widetilde{E}_r(f)=rE_0\frac{4\pi ifx(f)}{c}\,.
\end{equation}

Here, $x(f)$ represents the spectral characteristics of the mirror's position.

It is noteworthy that, in the absence of a beam splitter, the optical configuration illustrated in Figure 3 effectively manifests itself as a system featuring two inputs and two outputs. In our subsequent calculations, we delve into the AC spectra of the output fields generated by light input through both the upper and lower paths of the central vortex. The DC field originating from either the upper or lower path contributes to AC spectra at both output paths, and each output is responsive to the motion of ETMx or ETMy. Alternatively, this translates to sensitivity to the noise spectra of $L_x$ and $L_y$. Consequently, we embark on the computation of the transfer functions for all possible combinations—eight in total—of field components. Ultimately, we derive the final transfer functions at the output by means of linear combinations. The merits of this approach lie in its straightforward derivation of each of these eight components.

For simplicity, in the following derivation, we will use '1' to denote the upper path and '2' for the lower path in the forthcoming derivations. Furthermore, a dual subscript notation will be employed to designate the AC spectral characteristics of the output fields. In this notation, the first number indicates the input path of the original DC field, while the second number indicates the output path for the field spectra. To illustrate, $\widetilde{E}_{11}$ signifies the AC spectrum of the output field at the upper path of the central vortex, generated by the quasi-static field component injected into the L-shaped resonator from the upper path.

Firstly, we consider the scenario involving the single-side injection of a DC field from the upper path. We represent the static input component at frequency $f_0$ as $E_{in,1}$. Next, we calculate the static field components at the end mirrors, which are given by:

\begin{equation}\label{C_2}
\overline E_{0,ETMx}=\frac{t_i}{1-r_i^2}e^{i\Psi_1}E_{in,1}\,,
\end{equation}

\begin{equation}\label{C_3}
\overline E_{0,ETMy}=\frac{t_ir_i}{1-r_i^2}e^{i\Psi_1}E_{in,1}\,,
\end{equation}

where $t_i$ and $r_i$ represent the transmissivity and reflectivity of the input test mass (ITM), respectively. $\Psi_1$ is defined as $\frac{2\pi if_0(l_1+l_2+L_x)}{c}$, accounting for the phase shift incurred during propagation along one of the shortest paths in the backward propagation mode. It's worth noting that, owing to the vertical arms of the L-shaped resonator being set to integer multiples of the laser wavelength, the round-trip propagation phase factor is omitted. The motion of the two end test masses introduces AC sidebands to the field components. These sideband spectra are further amplified in the L-shaped resonator before exiting the system. With this understanding, we can analytically compute the transfer functions for $L_x$ and $L_y$ at the two readout paths. For instance, the transfer function of the $L_x$ field at the upper output path can be expressed as:

\begin{equation}\label{C_5}
\frac{\partial \widetilde E_{11}}{\partial L_x}(f)=\frac{4\pi if_0}{c}\frac{t_i}{1-r_i^2e^{2i\phi}}e^{i(\Psi_1+\psi_1)}\overline E_{0,ETMx}=\frac{4\pi if_0}{c} \frac{1}{1-r_i^2e^{2i\phi}}e^{i(2\Psi_1+\psi_1)}E_{in,1}\,.
\end{equation}

Here, $\phi=\frac{4\pi ifL_+}{c}$ denotes the average phase delay of round-trip propagation in an arm of the L-shaped resonator 

The $L_y$ transfer function at the upper output path follows a similar pattern. The only distinction lies in the fact that the sideband signal produced at ETMy travels an additional distance of $L_x+L_y$ and undergoes an extra reflection at ITM before reaching the upper output port. The derivation of this transfer function is equally straightforward:

\begin{equation}\label{C_5_1}
\frac{\partial \widetilde E_{11}}{\partial L_y}(f)=\frac{4\pi if_0}{c}\frac{r_it_i}{1-r_i^2e^{2i\phi}}e^{i(\Psi_1+\psi_1)}e^{i\phi}\overline E_{0,ETMy}=\frac{4\pi if_0}{c} \frac{r_i^2}{1-r_i^2e^{2i\phi}}e^{i(2\Psi_1+\psi_1)}e^{i\phi}E_{in,1}\,,
\end{equation}

where $\Psi_2=\frac{2\pi if_0(l_3+l_4+L_y)}{c}$ and $\psi_2=\frac{2\pi if(l_3+l_4+L_y)}{c}$ is the propagation phase of carrier and sideband signal along the other shortest path.

Similarly, we could get the transfer functions at the lower output port:

\begin{equation}\label{C_6}
\frac{\partial \widetilde E_{12}}{\partial L_x}(f)=\frac{4\pi if_0}{c}\frac{r_it_i}{1-r_i^2e^{2i\phi}}e^{i(\Psi_2+\psi_2)}e^{i\phi}\overline E_{0,ETMx}=\frac{4\pi if_0}{c} \frac{r_i}{1-r_i^2e^{2i\phi}}e^{i(\Psi_1+\Psi_2+\psi_2)}e^{i\phi}E_{in,1}\,,
\end{equation}

\begin{equation}\label{C_7}
\frac{\partial \widetilde E_{12}}{\partial L_y}(f)=\frac{4\pi if_0}{c}\frac{t_i}{1-r_i^2e^{2i\phi}}e^{i(\Psi_2+\psi_2)}\overline E_{0,ETMy}=\frac{4\pi if_0}{c} \frac{r_i}{1-r_i^2e^{2i\phi}}e^{i(\Psi_1+\Psi_2+\psi_2)}E_{in,1}\,,
\end{equation}

Following the same steps, we could calculate the relative transfer functions with single-side injection from the lower path:

\begin{equation}\label{C_8}
\frac{\partial \widetilde E_{22}}{\partial L_x}(f)=\frac{4\pi if_0}{c} \frac{r_i^2}{1-r_i^2e^{2i\phi}}e^{i(2\Psi_2+\psi_2)}e^{i\phi}E_{in,2}\,,
\end{equation}

\begin{equation}\label{C_9}
\frac{\partial \widetilde E_{22}}{\partial L_y}(f)=\frac{4\pi if_0}{c} \frac{1}{1-r_i^2e^{2i\phi}}e^{i(2\Psi_2+\psi_2)}E_{in,2}\,,
\end{equation}

\begin{equation}\label{C_10}
\frac{\partial \widetilde E_{21}}{\partial L_x}(f)=\frac{4\pi if_0}{c} \frac{r_i}{1-r_i^2e^{2i\phi}}e^{i(\Psi_1+\psi_1+\Psi_2)}E_{in,2}\,,
\end{equation}

\begin{equation}\label{C_11}
\frac{\partial \widetilde E_{21}}{\partial L_y}(f)=\frac{4\pi if_0}{c} \frac{r_i}{1-r_i^2e^{2i\phi}}e^{i(\Psi_1+\psi_1+\Psi_2)}e^{i\phi}E_{in,2}\,.
\end{equation}

The next step involves consolidating these findings to derive the final transfer function. Leveraging the relationship $E_{in,1}=E_{in,2}=\frac{E_{in}}{\sqrt{2}}$, we can obtain the $L_x$ and $L_y$ transfer functions at the bright and dark ports in Fig. 3. For instance, the $L_x$ transfer function at the bright port takes the following form:

\begin{equation}\label{C_12}
\frac{\partial E_{out1}}{\partial L_x}=\frac{1}{\sqrt{2}}\frac{\partial (E_{11}+E_{12}+E_{21}+E_{22})}{\partial L_x}=\frac{4\pi if_0}{c} \frac{(e^{i\Psi_1}+r_ie^{i\Psi_2})(e^{i(\Psi_1+\psi_1)}+r_ie^{i(\Psi_2+\psi_2)}e^{i\phi})}{1-r_i^2e^{2i\phi}}E_{in}\,.
\end{equation}
$$$$

For low-frequency sidebands, where $l_1$ and $l_2$ are much smaller than $L_+$, the phase factor $\psi_1$ can be approximated as $\phi/2$, as can $\psi_2$. Consequently, the transfer function can be simplified as:

\begin{equation}\label{C_13}
\frac{\partial E_{out1}}{\partial L_x}=\frac{1}{\sqrt{2}}\frac{\partial (E_{11}+E_{12}+E_{21}+E_{22})}{\partial L_x}=\frac{4\pi if_0}{c} \frac{(e^{i\frac{\delta \Psi}{2}}+r_ie^{-i \frac{\delta \Psi}{2}}e^{i\phi})(e^{i\frac{\delta \Psi}{2}}+r_ie^{-i \frac{\delta \Psi}{2}})}{1-r_i^2e^{2i\phi}}e^{i\Psi}e^{\frac{i\phi}{2}}E_{in}\,,
\end{equation}

where $\Psi=\frac{\Psi_1+\Psi_2}{2}$ is an overall phase factor, and $\delta \Phi=\frac{2\pi f_0(L_-+l_-)}{c}$ depicts the phase difference of two shortest path in the system.

Similarly, we could get the expression of the other transfer functions:

\begin{equation}\label{C_14}
\frac{\partial E_{out1}}{\partial L_y}=\frac{1}{\sqrt{2}}\frac{\partial (E_{11}+E_{12}+E_{21}+E_{22})}{\partial L_y}=\frac{4\pi if_0}{c} \frac{(e^{i\frac{-\delta \Psi}{2}}+r_ie^{i \frac{\delta \Psi}{2}}e^{i\phi})(e^{i\frac{-\delta \Psi}{2}}+r_ie^{i \frac{\delta \Psi}{2}})}{1-r_i^2e^{2i\phi}}e^{i\Psi}e^{\frac{i\phi}{2}}E_{in}\,,
\end{equation}

\begin{equation}\label{C_15}
\frac{\partial E_{out2}}{\partial L_x}=\frac{1}{\sqrt{2}}\frac{\partial (E_{11}+E_{21}-E_{12}-E_{22})}{\partial L_x}=\frac{4\pi if_0}{c} \frac{(e^{i\frac{\delta \Psi}{2}}-r_ie^{-i \frac{\delta \Psi}{2}}e^{i\phi})(e^{i\frac{\delta \Psi}{2}}+r_ie^{-i \frac{\delta \Psi}{2}})}{1-r_i^2e^{2i\phi}}e^{i\Psi}e^{\frac{i\phi}{2}}E_{in}\,,
\end{equation}

\begin{equation}\label{C_16}
\frac{\partial E_{out2}}{\partial L_y}=\frac{1}{\sqrt{2}}\frac{\partial (E_{11}+E_{21}-E_{12}-E_{22})}{\partial L_y}=-\frac{4\pi if_0}{c} \frac{(e^{-i\frac{\delta \Psi}{2}}-r_ie^{i \frac{\delta \Psi}{2}}e^{i\phi})(e^{-i\frac{\delta \Psi}{2}}+r_ie^{i \frac{\delta \Psi}{2}})}{1-r_i^2e^{2i\phi}}e^{i\Psi}e^{\frac{i\phi}{2}}E_{in}\,.
\end{equation}

When the $L_- + l_-$ degree of freedom and the common degrees of freedom are basically locked， it is evident that the $L_x$ and $L_y$ transfer functions exhibit complete symmetry at the bright port and antisymmetry at the dark port. This indicates that the bright port is sensitive to fluctuations in the common degree of freedom, $L_+$, while the dark port is sensitive to fluctuations in the differential degree of freedom, $L_-$. Through a straightforward linear combination, we can obtain the two required transfer functions:

\begin{equation}\label{eq12345_1}
{\cal T}(L_+) \equiv \frac{1}{2} (\frac{\partial E_{out1}}{\partial L_x}+\frac{\partial E_{out1}}{\partial L_y}) =  \frac{4\pi iE_{\rm in}}{\lambda}\frac{[r_{\rm i}(1+\mathrm{e}^{i\phi})+(1+r_{\rm i}^2\mathrm{e}^{i\phi})\mathrm{cos}(\delta \Psi)]\mathrm{e}^{\frac{i\phi}{2}}}{1-r^2_{\rm i}\mathrm{e}^{2i\phi}}\,,
\end{equation}

\begin{equation}\label{eq123456_1}
{\cal T}(L_-) \equiv \frac{1}{2} (\frac{\partial E_{out2}}{\partial L_x}-\frac{\partial E_{out2}}{\partial L_y}) =  \frac{4\pi iE_{\rm in}}{\lambda}\frac{[r_{\rm i}(1-\mathrm{e}^{i\phi})+(1-r_{\rm i}^2\mathrm{e}^{i\phi})\mathrm{cos}(\delta \Psi)]\mathrm{e}^{\frac{i\phi}{2}}}{1-r^2_{\rm i}\mathrm{e}^{2i\phi}}\,.
\end{equation}

These results are consistent with equations (6) and (7). Following the same procedure, one can also directly derive the $l_-$ transfer function in the system, as expressed in equation (8).

\section*{References}

\bibliographystyle{unsrt}
\bibliography{reference}
\end{CJK}
\end{document}